\documentstyle[prd,aps,epsf,floats]{revtex} 
\draft

\begin{document}
\twocolumn[\hsize\textwidth\columnwidth\hsize\csname
@twocolumnfalse\endcsname
\title{
\hbox to\hsize{\large Submitted to Phys.~Rev.~D \hfil E-Print
astro-ph/9703056}
\vskip1.55cm
Nucleon Spin Fluctuations and Neutrino-Nucleon
Energy Transfer in Supernovae}
\author{G\"unter Sigl$^*$}
\address{Max-Planck-Institut f\"ur Physik, F\"ohringer Ring 6,
D-80805 M\"unchen, Germany}
\date{\today}
\maketitle
\begin{abstract}
The formation of neutrino spectra in a supernova depends crucially
on strength and inelasticity of weak interactions in hot nuclear
matter. Neutrino interactions with nonrelativistic nucleons
are mainly governed by the dynamical structure function for the
nucleon spin density which describes its fluctuations. It has
recently been shown that these fluctuations give rise to a new
mode of energy transfer
between neutrinos and nucleons which inside the neutrinosphere
is of comparable or greater importance than ordinary recoil.
We calculate numerically the spin density structure function
in the limit of a dilute, non-degenerate medium from exact
two-nucleon wave functions for some representative nuclear interaction
potentials. We show that spectrum and magnitude of the energy transfer
can deviate significantly from those based on the Born
approximation. They are, however, rather insensitive to the particular
nuclear potential as long as it reproduces experimental nucleon
scattering phase shifts at energies up to a few tens of MeV. We
also compare with calculations based on a one-pion exchange
potential in Born approximation and briefly comment on their
applicability near the center of a supernova core. Our study is
relevant for numerical simulations of the neutrino spectra
emerging from type-II supernovae.

\end{abstract}
\vskip2.2pc]


\narrowtext

\section{Introduction}

The detection of roughly a dozen neutrinos from SN 1987 A is in
good qualitative agreement with the neutrino signal expected from
the early cooling phase of a hot neutron star born in the center of the
collapsed core of a massive star~\cite{ST}. It is therefore generally
believed that type-II supernovae such as SN 1987 A are the optical
counterparts of such catastrophic events.

The formation of the spectra of neutrinos emitted from a type-II
supernova takes place in a region where weak neutral
current scattering and pair processes involving electron, $\mu$, and
$\tau$ neutrinos and charged current creation and absorption of electron
neutrinos on nucleons, nuclei and electrons cease to be efficient in keeping
the neutrinos in thermodynamical equilibrium with the medium. The
interplay between (roughly) energy conserving scattering and
energy changing reactions plays a crucial role in that respect~\cite{Janka}.
In previous studies of neutrino transport, the lowest order
neutrino opacities in vacuum have been used. Neutral current scattering
processes on nucleons and nuclei have been approximated to be
elastic~\cite{studies}. As a result, whereas the energy fluxes
predicted for the three neutrino flavors turn out to be very
similar~\cite{Janka}, the effective temperatures are significantly
higher for $\mu$ and $\tau$ neutrinos compared to electron neutrinos which
because of their more efficient energy exchange with the
medium decouple from it further out.

However, weak interaction rates in a medium differ significantly from those
taking place in vacuum. On the one hand, the spin-dependent strong
force between nucleons will establish spatial correlations of
the density and the spin-density in the medium which can either enhance
or reduce average interaction rates. Many papers on weak interactions
in neutron stars investigated these effects. However, they either
applied the Landau theory of quasiparticles assuming
a ``cold'' nuclear medium~\cite{Iwamoto1,cold} or the authors
focused on quasielastic scattering~\cite{Iwamoto1,Sawyer1,hot}
for which the energy transfer to the nucleons is smaller than
the momentum transfer.

On the other hand, a weak interaction
transferring an energy $\omega$ to the medium is sensitive to the
fluctuation power in density and spin-density at that frequency.
For example, at finite density, the spin-dependent nucleon-nucleon
interaction also causes the nucleon spins to fluctuate. This leads
to a reduction of the average total axial-vector current neutrino
scattering cross section compared to its vacuum value~\cite{Sawyer2,RSS}.
This effect is most important at the high temperatures pertaining in
the first few seconds after formation of the hot neutron star.
In addition, the nucleon spin fluctuations can, apart from recoil,
imply an enhanced energy transfer between nucleons and
neutrinos~\cite{JKRS}. It is this effect which we are mostly concerned
with in the present work because it
could significantly change predictions of the
neutrino spectra with a tendency to lower predicted effective temperatures
of $\mu$ and $\tau$ neutrinos~\cite{Janka}. This is of some
importance in view of new neutrino detectors such as Super-Kamiokande
and the Sudbury Neutrino Observatory, which have the capability
of measuring the neutrino spectra from nearby supernova events
with much better statistics than is available with the data from
SN 1987 A.

Near the surface of last scattering the neutrino opacities are
governed mainly by the local nucleon spin density. Within linear
response theory, weak neutral-current interactions are then determined
by the dynamical
structure function for the nucleon spin-density which describes its spatial
and temporal correlations and is a function of energy and momentum transfer.
For energy transfers that are larger than
the typical spin fluctuation rate multiple nucleon-nucleon scattering
is negligible, and, to lowest order in the nucleon-nucleon interaction,
the spin-density structure function can be calculated
from the matrix element for nucleon bremsstrahlung. This matrix
element has been discussed in some detail in the
literature~\cite{FM,BT,RS1} for a one-pion exchange (OPE)
potential to lowest order in the
pion-nucleon coupling, i.e. in Born approximation.

However, the Born approximation is only applicable if at least one of
the following conditions is fulfilled~\cite{Schiff}:
\begin{eqnarray}
  |V|&\ll&\frac{1}{m_N a^2}\nonumber\\
  |V|&\ll&\frac{p}{m_N a}\,,\label{Borncond}
\end{eqnarray}
where $|V|\sim100\,$MeV is the typical magnitude of the nuclear interaction
potential, $a\sim1\,$fm is its range, $p$ is the nucleon momentum in
the center of mass system, and $m_N$ is the free nucleon mass.
The first condition is always
violated if the potential leads to bound states as for
the proton-neutron interaction which gives rise to the
deuteron bound state. The second condition translates into
$p\gtrsim m_N$ and is therefore also violated for the non-relativistic
nucleon momenta occurring in a supernova.
We can therefore not expect that the Born approximation is
a reliable approximation to the dynamical nucleon spin-density
structure function in a supernova. Neither is it obvious that any
weak interaction rates calculated from it are reliable
at the relatively low energies involved.

The goal of this paper is therefore to compute the dynamical nucleon
spin-density structure function and resulting weak interaction rates
beyond the Born approximation by using exact two-nucleon wave functions. To
keep things numerically simple, we will restrict ourselves to
spherically symmetric but spin-dependent two-nucleon potentials.
Since a central
potential conserves the total nucleon spin, the only contribution
to inelastic weak processes (i.e. inelastic in the center of mass frame
of the nucleons) will then arise from interactions of protons
and neutrons due to their different weak coupling constants.
We therefore have to deal with two nucleon species.
Our approach takes into account in a consistent, unified way the
free-free transitions
\begin{equation}
  \nu+n+p\leftrightarrow\nu+n+p\label{free}
\end{equation}
as well as the free-bound and bound-free processes involving the
deuteron,
\begin{equation}
  \nu+d\leftrightarrow\nu+p+n\,.\label{deut}
\end{equation}
The analogous processes involving neutrino pairs or axions instead
of neutrino scattering are described by the same dynamical structure
function and can therefore also be treated by our formalism.

The rest of the paper is organized as follows: In Sect.~II we
define the nucleon spin-density structure function in a form
suitable for the case of proton-neutron interactions. The main
formalism for computing this structure function from two-nucleon
wave functions is presented in Sect.~III. In Sect.~IV the Born
approximation is derived as a limiting case. Sect.~V introduces
a central potential which reproduces experimental data on proton-neutron
scattering at energies below a few tens of MeV. The corresponding
spin-density structure function is
numerically calculated for conditions around the neutrino sphere
and compared with the Born approximation and calculations for
an OPE potential. Finally, we summarize and conclude in Sect.~VI.

\section{The Nucleon Spin-Density Structure Function}

\subsection{Definition and General Properties}

The interaction of interest here, neutrino-nucleon neutral-current
scattering, is given by the Hamiltonian
\begin{eqnarray}
  {\cal H}_{\rm int}&=&\frac{G_F}{2\sqrt2}\sum_{i=n,p}
  \bar\psi_i\gamma_\mu\left[C_{V,i}-C_{A,i}\gamma_5\right]\psi_i
  \nonumber\\
  &&\times\bar\psi_\nu\gamma^\mu(1-\gamma_5)\psi_\nu
  \,,\label{L1}
\end{eqnarray}
where $G_F$ is the Fermi constant, $\psi_i$ ($i=n,p$) and $\psi_\nu$ are
the Dirac field operators for the neutrons, protons and neutrinos, and
$C_{V,i}$ and $C_{A,i}$ are the dimensionless weak neutral-current vector and
axial-vector coupling constants for protons and neutrons, respectively.

Another possible type of weak process is the emission of
axions~\cite{Raffelt1}. The corresponding interaction Hamiltonian
reads
\begin{equation}
  {\cal H}_{\rm int}=\frac{1}{2f_a}\sum_{i=n,p}
  C_{a,i}\bar\psi_i\gamma_\mu\gamma_5\psi_i\partial^\mu a
  \,,\label{L2}
\end{equation}
where $a$ is the axion field, $f_a$ the Peccei-Quinn scale, and
the dimensionless coupling constants to neutrons and protons,
$C_{a,i}$ ($i=n,p$), are of order unity and depend on the specific
axion model.

In the limit of non-relativistic nucleons, only the axial-vector
coupling contributes to inelastic weak processes. Within linear
response theory these processes are then described exclusively by the
dynamical structure function for the nucleon spin density.
In the following, we will drop the index $A$ or $a$ in the nucleon
coupling constants to neutrinos and axions, respectively, for notational
simplicity whenever the result applies to both cases.
To ensure a suitable normalization which will become clear below
in Eq.~(\ref{norm}), we define the structure function as the
autocorrelation function of the weighted nucleon spin-density
\begin{equation}
  \hbox{\boldmath$\sigma$}_w(x)\equiv
  \sum_{i=n,p}\frac{C_i}{C}\phi_i^\dagger(x)\frac{\hbox{\boldmath$\tau$}}{2}
  \phi_i(x)\,.\label{sigmaw}
\end{equation}
Here, $\phi_i(x)$ ($i=n,p$) is the non-relativistic field operator
for protons and neutrons which is a Pauli two-spinor,
$\hbox{\boldmath$\tau$}$ is the vector of Pauli matrices,
and $C^2=C_n^2Y_n+C_p^2Y_p$ is an average neutral-current axial
weak coupling constant to the nucleons, weighted by the
fractional neutron and proton abundances $Y_n$ and $Y_p$.
Defining the Fourier transform in a
normalization volume $V$ as $\hbox{\boldmath$\sigma$}_w(t,{\bf k})=V^{-1/2}
\int d^3{\bf r}\,e^{-i{\bf k\cdot r}}\hbox{\boldmath$\sigma$}_w(t,{\bf r})$,
one can then define the nuceon spin-density structure
function~\cite{RS1,JKRS,Sigl}:
\begin{equation}
  S_\sigma(\omega,{\bf k})=
  \frac{4}{3n_b}\int_{-\infty}^{+\infty} dt\,
  e^{i\omega t}\langle\hbox{\boldmath$\sigma$}_w(t,{\bf k})\cdot
  \hbox{\boldmath$\sigma$}_w(0,{\bf -k})\rangle\,.\label{S1}
\end{equation}
Here, $(\omega,{\bf k})$ is the four-momentum transfer to the
medium, $n_b$ is the baryon number
density and the expectation value $\left\langle\cdots\right\rangle$ is
taken over a thermal ensemble at the medium temperature $T$ of medium
states normalized to unity.

Relativistic neutrinos and possibly axions will have
typical energies of order $3T$ but are in general not in chemical
equilibrium with the medium. Weak interactions
such as neutral current neutrino scattering and pair processes
and axion emission thus probe the spin-density function typically
at thermal energy-momentum transfers.
Since the momenta involved in the nucleon-nucleon interactions
are much larger than the thermal momenta of relativistic particles,
we will often employ the long wavelength limit,
$S_\sigma(\omega)\equiv\lim_{{\bf k}\to0}S_\sigma(\omega,{\bf k})$
for which we define the dimensionless quantity $\tilde{S}_\sigma(x)\equiv
TS_\sigma(xT)$. In this limit, integration of Eq.~(\ref{S1}) over
$\omega$ yields the sum rule
\begin{equation}
  \int_{-\infty}^{+\infty}\frac{d\omega}{2\pi}S_\sigma(\omega)-1
  =N_\sigma\equiv\frac{4}{3n_b V}\left\langle
  \sum_{i\neq j}\hbox{\boldmath$\sigma$}_{i,w}\cdot
  \hbox{\boldmath$\sigma$}_{j,w}
  \right\rangle\,,\label{norm}
\end{equation}
where we wrote the spatial integral
$\int d^3{\bf r}\,\hbox{\boldmath$\sigma$}_w(t,{\bf r})
=\sum_i\hbox{\boldmath$\sigma$}_{i,w}$. Here,
$\hbox{\boldmath$\sigma$}_{i,w}\equiv\hbox{\boldmath$\sigma$}_i
\,{\rm diag}(C_p,C_n)/C$,
where $\hbox{\boldmath$\sigma$}_i$ are the spin operators of the
individual nucleons, and the matrix ${\rm diag}(C_p,C_n)$
acts in isospin space. In Eq.~(\ref{norm}) $N_\sigma$ describes
correlations among different nucleon spins. In the absence of
such correlations $\int_{-\infty}^{+\infty}(d\omega/2\pi)
S_\sigma(\omega)$ reduces to 1 which motivated the introduction
of the weighted spin operator Eq.~(\ref{sigmaw}).

We formally introduce the complete set of eigenfunctions
$\left|n\right\rangle$ of the total Hamiltonian $H$ of the nuclear medium,
$H\left|n\right\rangle=\omega_n\left|n\right\rangle$, where $\omega_n$ are
the corresponding energy eigenvalues. By inserting the identity
operator $I=\left|n\right\rangle\left\langle n\right|$, between the
spin operators, Eq.~(\ref{S1}) can be rewritten into
\begin{eqnarray}
  S_\sigma(\omega,{\bf k})&=&\frac{8\pi}{3n_b}\frac{1}{Z}
  \sum_{n,m}e^{-\omega_n/T}\delta(\omega+\omega_n-\omega_m)
  \nonumber\\
  &&\hskip1.5cm\times\left|\left\langle n|
  \hbox{\boldmath$\sigma$}_w(0,{\bf k})|m\right\rangle\right|^2
  \,,\label{S2}
\end{eqnarray}
where $Z=\sum_n e^{-\omega_n/T}$ is the partition function. This
form will be useful later and it shows that the structure function
satisfies detailed balance,
\begin{equation}
  S_\sigma(\omega,{\bf k})=S_\sigma(-\omega,-{\bf k})e^{\omega/T}
  \,.\label{detbal}
\end{equation}
It is therefore sufficient to know the function, e.g. for positive
energy transfer to the medium, $\omega\geq0$.

Up to now
no approximations have been made with regard to the nucleon-nucleon
interactions which determine the nonperturbative though unknown
structure function $S_\sigma(\omega,{\bf k})$. From now on we
will make the assumption that only two-nucleon forces are present.
The Hamiltonian then has the form
\begin{equation}
  H=\sum_i\frac{{\bf p}_i^2}{2m_N}+\frac{1}{2}
  \sum_{i\neq j}V({\bf r}_{ij},{\hbox{\boldmath $\sigma$}}_i,
  {\hbox{\boldmath $\sigma$}}_j)\,,\label{H}
\end{equation}
where ${\bf r}_{ij}$ is the radius vector between nucleon $i$ and
$j$, ${\bf p}_i$ is the nucleon momentum,
$V({\bf r}_{ij},{\hbox{\boldmath$\sigma$}}_i,
{\hbox{\boldmath $\sigma$}}_j)$ is the spin
dependent two-nucleon interaction potential, and the sums run over
all nucleons. The most general
two-nucleon potential in the non-relativistic limit can be written
as~\cite{MS}
\begin{eqnarray}
  V({\bf r},{\hbox{\boldmath $\sigma$}}_1,{\hbox{\boldmath
  $\sigma$}}_2)&=&U(r)+U_\sigma(r){\hbox{\boldmath $\sigma$}}_1
  \cdot{\hbox{\boldmath $\sigma$}}_2
  +U_T(r)T_{12}\label{Vint}\\
  &&+P_\tau\left[U^\tau(r)+U_\sigma^\tau(r){\hbox{\boldmath $\sigma$}}_1
  \cdot{\hbox{\boldmath $\sigma$}}_2
  +U_T^\tau(r)T_{12}\right]\nonumber
\end{eqnarray}
where ${\bf r}={\bf r}_{12}$, $r=\vert{\bf r}\vert$, $\hat{\bf r}=
{\bf r}/r$, $P_\tau$ is the isospin exchange operator, and the tensor
operator is given by
\begin{equation}
  T_{12}=3\,{\hbox{\boldmath$\sigma$}}_1\cdot\hat{\bf r}\,
  {\hbox{\boldmath $\sigma$}}_2\cdot\hat{\bf r}-
  {\hbox{\boldmath $\sigma$}}_1
  \cdot{\hbox{\boldmath $\sigma$}}_2\,.\label{tensor}
\end{equation}

Useful information about structure functions is contained
in their moments of which Eq.(\ref{norm}) is an example for the lowest one.
The next higher moment is given by the so called f-sum rule which is often
discussed in the literature in the context of the density structure
function and for spin-conserving interactions~\cite{Forster}.
In Ref.~\cite{Sigl} we derived a generalized
f-sum rule for the spin-density structure function for one species
of nucleons interacting via spin-dependent forces of the form
Eq.~(\ref{Vint}) in a non-degenerate medium:
\begin{equation}
  \int_{-\infty}^{+\infty}\frac{d\omega}{2\pi}\,\omega
  S_\sigma(\omega)=-\frac{4}{n_b V}\left\langle H_T\right\rangle
  \,.\label{fsum1}
\end{equation}
Here, $H_T$ is the part
of the total Hamiltonian involving the tensor operator $T_{ij}$.
In the present paper we will consider both neutrons and protons but
assume a central two-nucleon potential, i.e. absence of tensor
contributions. The f-sum rule is then modified to
\begin{equation}
  \int_{-\infty}^{+\infty}\frac{d\omega}{2\pi}\,\omega
  S_\sigma(\omega)=-\frac{4}{3n_b V}
  \left(\frac{C_p-C_n}{C}\right)^2
  \left\langle H_\sigma^{np}\right\rangle
  \,,\label{fsum2}
\end{equation}
where $H_\sigma^{np}$ is the spin-dependent central part of the total
Hamiltonian which contributes to neutron-proton interactions. Note from
Eqs.~(\ref{detbal}) and (\ref{fsum1}) that for
only one nucleon species a tensor interaction is required to give
a non-trivial spin-density structure function. This is because the
central part of the interaction conserves the total nucleon spin
and thus does not contribute to its fluctuations.
In contrast, for two nucleon species, a central spin-dependent
proton-neutron interaction
is sufficient for a non-trivial structure function as long
as the neutral-current axial weak coupling constants 
for protons and neutrons are different [see Eq.~(\ref{fsum2})].
We stress, however, that
the actual (positive) value of the f sum depends on all interaction terms
via the states entering the thermal average.

\subsection{Relevance for Weak Interactions}

The differential axial-vector-current neutrino-nucleon cross section is
determined by the dynamical nucleon spin-density structure function
$S_\sigma(\omega,{\bf k})$, taken at the difference of initial and
final neutrino four-momentum $(\varepsilon_1,{\bf k}_1)$ and
$(\varepsilon_2,{\bf k}_2)$ via~\cite{RS1,Sigl}:
\begin{equation}
  d\sigma_A=G_F^2C_A^2\frac{3-\cos\theta}{4}
  S_\sigma(\varepsilon_1-\varepsilon_2,{\bf k}_1-{\bf k}_2)
  \frac{d^3{\bf k}_2}{(2\pi)^3}\,,\label{dsigma}
\end{equation}
where $\theta$ is the angle between ${\bf k}_1$ and ${\bf k}_2$.
In our convention, the neutral current axial-vector contribution
to neutrino scattering rates on the ensemble of all nucleons is
$n_b d\sigma_A$.

The axion emission rate per volume, $Q_a$, is governed by a
structure function $S_{\sigma,a}$ which is obtained from
Eqs.~(\ref{sigmaw}), (\ref{S1})
by substituting $C_i\to C_{a,i}$ ($i=n,p$),
\begin{equation}
  Q_a={C_a^2n_b\over(4\pi)^2f_a^2}\int_0^\infty d\omega\,\omega^4
  S_{\sigma,a}(-\omega,\omega)\,,\label{Qa}
\end{equation}
where $C_a^2=C_{a,n}^2Y_n+C_{a,p}^2Y_p$.
We have assumed an isotropic medium such that $S_\sigma(\omega,{\bf k})$
only depends on $k=|{\bf k}|$.

Various quantities relevant for neutrino diffusion are determined
by the spin-density structure function. For the remainder of
this section, we assume a
Maxwell-Boltzmann distribution at temperature $T_\nu$ for the neutrinos.
Furthermore, we make use of the normalization given by Eq.~(\ref{norm}).
Contributions form spin-spin correlations represented by $N_\sigma$
are mainly induced by the presence of nucleon bound states and
by the Pauli exclusion principle which becomes important in a
degenerate medium~\cite{Strobel}. Both effects are small in the
post-collapse phase of a supernova in which we are interested.
We therefore assume $N_\sigma\ll1$ in Eq.~(\ref{norm}).
The average energy transfer per collision in a dilute medium can then be
written as~\cite{JKRS,RSS}
\begin{equation}
  \frac{\langle\Delta\varepsilon\rangle}{T}
  =\int_0^\infty\frac{dx}{2\pi}\,\tilde{S}_\sigma(x)
  \left(x+\frac{\beta x^2}{2}+\frac{\beta^2x^3}{12}\right)
  (e^{-x}-e^{-\beta x})\,,\label{Etrans}
\end{equation}
with $\beta\equiv T/T_\nu$. This should be compared to the average
energy transfer by nucleon recoils~\cite{Tubbs},
\begin{equation}
  \langle\Delta\varepsilon\rangle_{\rm recoil}=\frac{30(\beta-1)}{\beta^2}
  \frac{T^2}{m_N}\,.\label{recoil}
\end{equation}
Another interesting quantity is the reduction of the average
total axial-vector current scattering cross section
$\langle\sigma_A\rangle$ [see Eq.~(\ref{dsigma})] in the nuclear
medium~\cite{Sawyer2,RSS}. First we note that
a term of the form $A\delta(\omega)$ in $S_\sigma(\omega)$ corresponds
to a total elastic scattering cross section
\begin{equation}
  \sigma_{el}(\varepsilon_1)=\frac{3A}{8\pi^2}G_F^2C_A^2
  \varepsilon_1^2\,.\label{elastic}
\end{equation}
For $n_b\to0$ there are no spin fluctuations and correlations, and
Eq.~(\ref{norm}) implies $S_\sigma(\omega)=2\pi\delta(\omega)$ and
thus $\sigma_0\equiv(9/\pi)C_A^2G_F^2T^2$ for the thermally averaged
cross section. In Ref.~\cite{RSS} we obtained the expression
\begin{eqnarray}
  \frac{\delta\langle\sigma_A\rangle}{\sigma_0}&\equiv&
  \frac{\langle\sigma_A\rangle-\sigma_0}{\sigma_0}\label{deltasigma}\\
  &=&N_\sigma-\int_0^\infty\frac{dx}{2\pi}\,\tilde{S}_\sigma(x)
  \left[1-\left(1+x+\frac{x^2}{6}\right)e^{-x}\right]
  \,,\nonumber
\end{eqnarray}
which again holds in the dilute medium. The physical quantities
discussed here will be calculated for the
supernova environment in Sect.~V.

\section{Beyond the Born Approximation}

\subsection{Classical versus General Quantum Result}

In the limit $|\omega|\ll T$ the nucleon spin can be treated as
a classical spin ${\bf s}$ being changed abruptly by some random
amount $\Delta{\bf s}$ in a typical nucleon-nucleon
collision event which takes place on a time scale $\simeq1/T$ and
thus appears to be ``hard''. In this case we expect~\cite{Raffelt}
\begin{equation}
  S_\sigma(\omega)\simeq\frac{\Gamma_\sigma}{\omega^2+\Gamma_\sigma^2/4}
  \,,\label{classical}
\end{equation}
where the spin fluctuation rate $\Gamma_\sigma$ is related to
the collision rate $\Gamma_{\rm coll}$ by
\begin{equation}
  \Gamma_\sigma=\frac{\left\langle(\Delta{\bf s})^2\right\rangle}
  {\langle{\bf s}^2\rangle}\,\Gamma_{\rm coll}\,.\label{Gsigma}
\end{equation}
Note that the spin fluctuation rate suppresses the $\omega^{-2}$
bremsstrahlung spectrum which otherwise would violate the
existence of the normalization Eq.~(\ref{norm}). This is known
as the Landau-Pomeranchuk-Migdal (LPM) effect~\cite{LPM,KV}.
In previous work~\cite{RS2,KJR,JKRS,Sigl} it has been discussed
how the Lorentzian
shape Eq.~(\ref{classical}) might influence weak interaction rates
at high densities where $\Gamma_\sigma\gtrsim T$. The high density
behavior of the spin-density structure function can also influence
limits on the axion mass~\cite{KJSSTE}.

For $|\omega|\gg\Gamma_\sigma$ multiple scattering effects can be ignored
and the spin-density structure function can be computed by a
quantum mechanical treatment of two-nucleon scattering. From the
generic $\omega^{-2}$ divergence of all bremsstrahlung processes
for $\omega\to0$ one expects the general form~\cite{RSS}
\begin{equation}
  S_\sigma(\omega)=\frac{\Gamma_\sigma}{\omega^2}\,s(\omega/T)
  \times\cases{e^{\omega/T}&for $\omega<0$,\cr
  \noalign{\vskip3pt plus1pt}
  1 &for $\omega>0$,\cr}\label{quantum}
\end{equation}
where $s(x)$ is a nonsingular even function with $s(0)=1$. The
specific shape of $s(x)$ for $x\gtrsim1$ depends on the nucleon-nucleon
interaction potential Eq.~(\ref{Vint}), and its calculation for
realistic interaction potentials is the main goal of this paper.
Comparing Eqs.~(\ref{classical}) and (\ref{quantum}) in their
common range of validity, $\Gamma_\sigma\ll|\omega|\ll T$, shows
that the coefficient $\Gamma_\sigma$ of the bremsstrahlung
divergence in Eq.~(\ref{quantum}) can be interpreted as a nucleon
spin fluctuation rate and
that the classical limit of hard collisions corresponds to
$s(x)=1$. The existence of the f sum Eqs.~(\ref{fsum1}), (\ref{fsum2})
shows that $s(x)$ has to decrease for large $x$ due to quantum
corrections.

\subsection{Exact Treatment in the Limit of High Energy Transfers}

For $\omega\gtrsim\Gamma_\sigma$ where
scattering involving more than two nucleons is negligible, we can
numerically compute two-nucleon wave functions from a given nucleon-nucleon
interaction potential and use it in the general expression
Eq.~(\ref{S2}). For a central potential the eigenfunctions for the
relative motion in the proton-neutron center of mass system
\begin{equation}
  |P\rangle\equiv|p,l,m,S\rangle=R_{plS}(r)Y_{lm}(\Omega)|S\rangle
\end{equation}
are characterized by the quantum numbers for the radial momentum,
$p$, the orbital angular momentum, $l$ and $m$, and the total spin, $S$,
where $R_{plS}(r)$ is the radial wave function and $Y_{lm}(\Omega)$ are the
spherical harmonics. The corresponding energy eigenvalues $\omega_P$
have a $(2l+1)(2S+1)$ fold degeneracy. The Pauli exclusion principle
then also determines the isospin to $I=\frac{1}{2}+(-1)^l\left(
\frac{1}{2}-S\right)$. Assuming an isotropic medium and using
$\hbox{\boldmath$\sigma$}_w(0,{\bf k})=V^{-1/2}\sum_i
\hbox{\boldmath$\sigma$}_ie^{-i{\bf k}\cdot{\bf r}_i}
\,{\rm diag}(C_p,C_n)/C$ with a normalization volume $V=1/n_b$,
after some algebraic manipulations we obtain
\begin{eqnarray}
  S_\sigma(\omega,k)&=&\frac{16\pi^{1/2}}{3C^2}\frac{1}{k}
  \left(\frac{m_N}{T}\right)^{1/2}Y_pY_n\frac{1}{Z_{CM}}\label{S3}\\
  &&\times\sum_{P,Q}
  e^{-\omega_P/T-m_N\left[\omega+\omega_P-\omega_Q-
  k^2/(4m_N)\right]^2/(Tk^2)}\nonumber\\
  &&\times\left|\left\langle P\left|
  C_p\hbox{\boldmath$\sigma$}_p e^{-i\frac{{\bf k}\cdot{\bf r}}{2}}+
  C_n\hbox{\boldmath$\sigma$}_n e^{+i\frac{{\bf k}\cdot{\bf r}}{2}}
  \right|Q\right\rangle\right|^2\,,\nonumber
\end{eqnarray}
where $Z_{CM}=\sum_{P}e^{-\omega_P/T}$.
For $k\to0$ this expression transforms into
\begin{eqnarray}
  S_\sigma(\omega)&=&4\pi\,Y_pY_n
  \left(\frac{C_p-C_n}{C}\right)^2\frac{1}{Z_{CM}}\label{S4}\\
  &&\times\sum_{p,q}\sum_l(2l+1)\sum_{S=0}^1 e^{-\omega_P/T}
  \delta(\omega+\omega_P-\omega_Q)\nonumber\\
  &&\times\left|\int_0^{r_{\rm max}}dr
  r^2R_{plS}^*(r)R_{ql(1-S)}(r)\right|^2\,,\nonumber
\end{eqnarray}
where we have made use of the orthogonality of the system of eigenfunctions
which are supposed to be normalized to unity. In practice one constructs
bound and scattering
states of the stationary radial Schr\"odinger equation within a finite
spherical volume of radius $r_{\rm max}$ and computes the matrix
elements appearing in Eq.~(\ref{S4}). For a nucleon interaction potential
that is not radially symmetric, the eigenstates
$|P\rangle\equiv|pJ{\cal P}\rangle$ are characterized by the total angular
momentum $J$ and parity ${\cal P}$ and are superpositions of
orbital angular momentum eigenstates. With this modification,
Eq.~(\ref{S3}) still holds but we will not pursue this more complicated
case here which would lead to coupled radial equations
for the corresponding radial functions $R_{pJ{\cal P}}$.

Since we neglect interactions among more than two nucleons, our
formalism does only account for neutrons, protons and deuterons.
Higher nuclei such as helium are not included. In this sense, strictly,
$n_b$, $Y_p$, and $Y_n$ have to be interpreted within the ensemble
of neutrons, protons and deuterons only. Nuclear statistical
equilibrium shows that in practice this does not make a big
difference in our situation where $Y_p\ll Y_n$. Keeping this in mind
we can now calculate $Z_{CM}$
analytically. Around the neutrinosphere the nucleons are at best mildly
degenerate. We therefore assume a Maxwell-Boltzmann distribution
$f(p)=e^{-p^2/(2\mu T)}$ for the unbound proton-neutron states. Here,
$p=s^{1/2}/2$ is the nucleon momentum in the center of mass system,
expressed in terms of the squared center of mass energy $s$
(excluding the nucleon rest mass),
and $\mu=m_N/2$ is the reduced nucleon mass.
Taking into account the spin degrees of freedom we then have
\begin{equation}
  Z_{CM}=3e^{-\varepsilon_d/T}+\frac{4}{n_b}
  \left(\frac{\mu T}{2\pi}\right)^{3/2}\,,\label{ZCM}
\end{equation}
where $\varepsilon_d\simeq2.2\,$MeV is the deuteron binding energy (the
deuteron has $S=1$). The degree of dissociation, i.e. the fractional
abundance of unbound states is then
\begin{equation}
  f_u=\left[1+\frac{3}{4}n_b\left(\frac{2\pi}{\mu T}\right)^{3/2}
  e^{-\varepsilon_d/T}\right]^{-1}\,.\label{fd}
\end{equation}
As a consequence, $S_\sigma(\omega)$ from Eqs.~(\ref{S3}) and (\ref{S4})
does not exhibit a simple linear scaling with the nucleon density
$n_b$, except for the dilute limit, $n_b\to0$, $f_b\to1$. It can
be seen that the numerator in Eqs.~(\ref{S3}) and (\ref{S4}) is independent
of $n_b$ and the density dependence of $S_\sigma$ thus exclusively
stems from $Z_{CM}$.

In the limit of zero temperature, $f_u\to0$ and only the deuteron bound state
will be populated in thermal equilibrium. In this limit, Eq.~(\ref{dsigma})
describes the cross section for the weak neutral current deuteron break up
process, Eq.~(\ref{deut}).
Integration over the phase space for the outgoing neutrino yields
the total cross section
\begin{eqnarray}
  \sigma_{\nu d}^{\rm NC}(\varepsilon_1)&=&\frac{3G_F^2}{16\pi^2}
  \frac{(C_p-C_n)^2}{Y}\label{sigmaD}\\
  &&\times\int_{\varepsilon_d}^{\varepsilon_1}d\omega
  (\varepsilon_1-\omega)^2
  \left[\lim_{T\to0}S_\sigma(\omega)\right]\nonumber
\end{eqnarray}
for incident neutrino energy $\varepsilon_1$.

Another instructive limiting case is the absence of spin flip
interactions. Scattering on protons and neutrons then has to be
added incoherently with the states $|P\rangle$
now being plane waves. Eq.~(\ref{S3}) then
reduces to the ordinary recoil expression
\begin{equation}
  S_\sigma(\omega,k)=\frac{1}{k}\left(\frac{2\pi m_N}{T}\right)^{1/2}
  e^{-m_N\left[\omega-k^2/(2m_N)\right]^2/(2Tk^2)}\,.\label{Srecoil}
\end{equation}

Let us now get back to the general case.
In agreement with Eq.~(\ref{fsum2}) and Ref.~\cite{FM} only proton-neutron
scattering contributes to Eq.~(\ref{S4}), and only if the
neutral-current axial weak
coupling constants for protons and neutrons are different. Since the
total spin is conserved by a central potential, the spin-density
structure function is governed by the fluctuations of the difference
of the proton and neutron spin,
$\hbox{\boldmath$\sigma$}_p-\hbox{\boldmath$\sigma$}_n$. To compare
with the general results Eqs.~(\ref{classical}) and (\ref{quantum}),
we study the corresponding spin-flip cross section which is defined as
\begin{equation}
  \sigma_{sf}(s)=\frac{\left\langle[\Delta
  (\hbox{\boldmath$\sigma$}_p-\hbox{\boldmath$\sigma$}_n)]^2
  \right\rangle}{\langle\left(\hbox{\boldmath$\sigma$}_p-
  \hbox{\boldmath$\sigma$}_n\right)^2\rangle}\,
  \sigma_{np}(s)\,.\label{sigmasf}
\end{equation}
Here, $\sigma_{np}(s)=\sum_l\sigma_{np}(l,s)$, where the average total
proton-neutron scattering cross section in angular momentum state
$l$ is given by
\begin{equation}
  \sigma_{np}(l,s)=\frac{4\pi}{s}(2l+1)
  \left[3\sin^2\delta_{l,1}(s)+\sin^2\delta_{l,0}(s)\right]
  \label{sigmanum}
\end{equation}
in terms of the phase shifts $\delta_{lS}(s)$. The latter are
defined by the asymptotic behavior
\begin{equation}
  R_{plS}(r)\propto\sin\left[(2\mu\omega_P)^{1/2}r-l\pi/2+
  \delta_{lS}(8\mu\omega_P)\right]\,,\label{phasedef}
\end{equation}
of the scattering states $\omega_P>0$ for $(2\mu\omega_P)^{1/2}r\gg l$.
The nucleon spin flip rate is now just defined as
\begin{equation}
  \Gamma_{sf}=Y_p Y_n n_b\frac{\int_0^{+\infty}dp p^2f(p)
  (p/\mu)\sigma_{sf}(4p^2)}
  {\int_0^{+\infty}dp p^2f(p)}\,,\label{Gammasf}
\end{equation}
where $p/\mu$ is the relative velocity of proton and neutron.

As can be seen from phase shift analysis,
the spin flip cross section Eq.~(\ref{sigmasf}) is
\begin{equation}
  \sigma_{sf}(s)=\frac{16\pi}{3s}\sum_l(2l+1)
  \left[\sin\delta_{l,1}(s)-\sin\delta_{l,0}(s)\right]^2
  \,.\label{sigmasfnum}
\end{equation}
Note that this vanishes if the
phase shifts for $S=0$ and $S=1$ are equal, as expected.
Given $\sigma_{sf}$, one can compute $\Gamma_{sf}$ from
Eq.~(\ref{Gammasf}) and compare it with the Born approximation
to be discussed below and with Eq.~(\ref{quantum}). This will
be done in the following two sections.

\section{The Born Approximation}

By expanding the unbound states $|P\rangle$ into plane waves within
first order perturbation theory and inserting the result into
Eq.~(\ref{S3}), one obtains the spin-density structure function in
Born approximation. For $\omega>0$, the result
in the long wavelength limit is
\begin{eqnarray}
  S_\sigma^{\rm Born}(\omega)&=&\frac{1}{\omega^2}\frac{\mu}{2\pi}
  Y_p Y_n n_b\left(\frac{C_p-C_n}{C}\right)^2
  \label{SBorn}\\
  &&\times\frac{\int_0^{+\infty}dp pf(p)\int_{k_{\rm min}}^{k_{\rm max}}
                dk k|U_\sigma^{np}(k)|^2}
               {\int_0^{+\infty}dp p^2f(p)}
  \,,\nonumber
\end{eqnarray}
where $k_{\rm max,min}=(p^2+2\mu\omega)^{1/2}\pm p$, and
$U_\sigma^{np}(k)$ is the Fourier transform of
the coefficient of $\hbox{\boldmath$\sigma$}_p\cdot\hbox{\boldmath
$\sigma$}_n$ in the proton-neutron interaction
($\hbox{\boldmath$\sigma$}_p$ and $\hbox{\boldmath$\sigma$}_n$ being
the proton and neutron spins).
Only the relative motion between proton and neutron influences
Eq.~(\ref{SBorn}) because neither energy nor momentum can be transferred
to the center of mass motion in the long wavelength limit. In
contrast to Eq.~(\ref{S4}), $S_\sigma^{\rm Born}(\omega)$ scales
linearly with $n_b$.

In Ref.~\cite{RSS} we considered one species of nucleons coupling
to a classical, external scattering center via an interaction of
the form Eq.~(\ref{Vint}) where one of the spins was replaced by
a classical spin ${\bf s}$ associated with the external scatterer.
The result for the spin-density structure function in the long
wavelength limit in Born approximation is very similar to
Eq.~(\ref{SBorn}) for the case of a central two-nucleon potential
and a medium of protons and neutrons with different neutral-current
axial weak coupling constants.

In Born approximation, the spin flip cross section Eq.~(\ref{sigmasf})
evaluates to
\begin{equation}
  \sigma_{sf}^{\rm Born}(s)=\frac{20}{27\pi}\frac{\mu^2}{s}
  \int_0^{s^{1/2}}dk k|U_\sigma^{np}(k)|^2\,,\label{sigmasfBorn}
\end{equation}
where $\langle\left(\hbox{\boldmath$\sigma$}_p-
\hbox{\boldmath$\sigma$}_n\right)^2\rangle=3/2$ was used.
Comparing Eqs.~(\ref{quantum}), (\ref{SBorn}), and (\ref{Gammasf})
yields
\begin{equation}
  \Gamma_\sigma^{\rm Born}=\frac{27}{10}\left(\frac{C_p-C_n}
  {C}\right)^2\,\Gamma_{sf}^{\rm Born}\,,\label{GammaBorn}
\end{equation}
i.e. the spin fluctuation rate in
$S_\sigma(\omega)$ and the average spin flip rate indeed agree
within a factor of order
unity, apart from the factor $[(C_p-C_n)/C]^2$ involving the weak
coupling constants which results from our specific definition of
$S_\sigma$.

As an example, we consider the usually adopted OPE
potential which is a good approximation to the nucleon-nucleon
interaction for distances greater than the inverse pion mass $m_\pi$.
With $f\simeq1$ being the pion-nucleon coupling constant, its Fourier
transform is
\begin{equation}
  V_{\rm OPE}({\bf k},{\hbox{\boldmath $\sigma$}}_1,
  {\hbox{\boldmath $\sigma$}}_2)=-\left({2f\over m_\pi}\right)^2
  {\left({\hbox{\boldmath $\sigma$}}_1\cdot{\bf k}\right)
  \left({\hbox{\boldmath $\sigma$}}_2\cdot{\bf k}\right)
  \over k^2+m^2_\pi}(2P_\tau-1)\label{OPE}
\end{equation}
and it clearly has a tensor contribution. The spin-density structure
function corresponding to this potential has been calculated in
Born approximation~\cite{RS1}. Translated
into our notation, the contribution from proton-neutron scattering takes
the form of Eq.~(\ref{quantum}) with $s(x)\equiv\tilde{s}(x)/
\tilde{s}(0)$ given by the function
\begin{eqnarray}
  \tilde{s}(x)&=&\int_0^\infty dv[v(v+x)]^{1/2}e^{-v}\label{tildes}\\
  &&\times\biggl[
  \left(5C_+^2+3C_-^2\right)s_1(v,x)+2\left(C_+^2+C_-^2\right)s_2(v,x)
  \nonumber\\
  &&\hskip0.5cm-\left(6C_+^2+2C_-^2\right)s_3(v,x)\biggr]\,,\nonumber
\end{eqnarray}
where $C_\pm=(C_p\pm C_n)/2$ and
\begin{equation}
  s_i(v,x)=\int_{-1}^{+1}dz
  \cases{\left(\frac{2v+x-2z[v(v+x)]^{1/2}}
  {2v+x-2z[v(v+x)]^{1/2}+y}\right)^2& $i=1$,\cr
  \noalign{\vskip3pt plus1pt}
  \frac{(2v+x)^2-4v(v+x)z^2}{(2v+x+y)^2-4v(v+x)z^2} &$i=2$,\cr
  \noalign{\vskip3pt plus1pt}
  \frac{x^2}{(2v+x+y)^2-4v(v+x)z^2} &$i=3$,\cr}\label{s123}
\end{equation}
with $y\equiv m_\pi^2/(m_N T)$. Furthermore, the contribution to the
nucleon spin fluctuation rate $\Gamma_\sigma$ is
\begin{equation}
  \Gamma_{\sigma, {\rm OPE}}^{\rm Born}=\frac{2}{3}Y_pY_n\,
  \frac{\tilde{s}(0)}{C^2}
  \Gamma_A\,,\label{GammaBornOPE}
\end{equation}
with
\begin{equation}
  \Gamma_A=4\sqrt\pi\,\alpha_\pi^2
  \frac{n_b T^{1/2}}{m_N^{5/2}}
  =8.6\,{\rm MeV}\,\rho_{13}\,T_{10}^{1/2}\,.\label{GammaA}
\end{equation}
Here, $\alpha_\pi\equiv(f 2m_N/m_\pi)^2/4\pi\approx15$,
$\rho_{13}\equiv\rho/10^{13}\,\rm g\,cm^{-3}$ with $\rho$ the density,
and $T_{10}\equiv T/10\,{\rm MeV}$. Note that
$S_{\sigma, {\rm OPE}}^{\rm Born}(\omega)\propto\omega^{-3/2}$ for
$\omega\to\infty$ and thus violates f sum integrability. As was explained
in Ref.~\cite{Sigl}, this is caused by the unphysical behavior of
the OPE potential for $r\to0$ that leads to an
$|V_{\rm OPE}({\bf k})|$ which for
$k\to\infty$ is asymptotically constant [see Eq.~(\ref{OPE})].

More generally, as can be seen from Eq.~(\ref{SBorn}), one has
$S_\sigma^{\rm Born}(\omega)\propto\omega^{-3/2-r}$ for
$\omega\to\infty$ if $|U_\sigma(k)|\propto k^{-r}$ for $k\to\infty$,
corresponding to existence and square integrability of the $(r-2)$th
derivative of the interaction potential. This should hold independently
of the Born approximation which is viable in the
limit of high energies as we will see now.

\section{A Numerical Model for the Supernova Environment}

We first note that $S_\sigma(\omega)$ from Eqs.~(\ref{SBorn}) and
(\ref{S3}) is proportional to the dimensionless factor
\begin{equation}
  Y\equiv Y_pY_n\left(\frac{C_p-C_n}{C}\right)^2=
  \left(C_p-C_n\right)^2\frac{Y_pY_n}{C_p^2Y_p+C_n^2Y_n}
  \label{fac}
\end{equation}
which describes its compositional and coupling constant dependence
for fixed $n_b$ and $T$.
For processes involving only protons or neutrons this factor
would be replaced by $Y_p^2$ and $Y_n^2$, respectively.
Since interaction rates are proportional to
$n_b S_\sigma$ by definition, $Y$ is a rough measure of the contribution
of proton-neutron scattering to weak neutral current inelastic
interaction rates. For the neutrino-nucleon coupling in a
nuclear medium we will adopt $C_{A,p}\simeq1.09$, and
$C_{A,n}\simeq-0.91$~\cite{RS1}, so that $Y\simeq0.5$ for $Y_p\simeq0.1$.

In the following we are interested in the environment given inside
but not far from the neutrinosphere in a supernova. For the rest of
this section we choose the representative
numbers $\rho=10^{13}\,{\rm gcm}^{-3}$, $T=8\,$MeV, and $T_\nu=10\,$MeV.
For these parameters, $f_u\simeq0.36$, corresponding to a fractional
deuterium abundance $Y_d\simeq(1-f_u)Y_p$. Nuclear statistical equilibrium
involving higher nuclei gives values that are within $20-30\%$ of this
if $Y_p\lesssim0.2$. We stress again that due to
the presence of bound states the spin-density
structure function calculated by phase shift analysis and weak
interaction rates computed from it do not exhibit a simple
scaling behavior with density and/or temperature, as discussed
below Eq.~(\ref{ZCM}). Our choice represents a typical case of
interest for the neutrino-nucleon energy transfer.

\begin{figure}[ht]
\centering\leavevmode
\epsfxsize=3.2in
\epsfbox{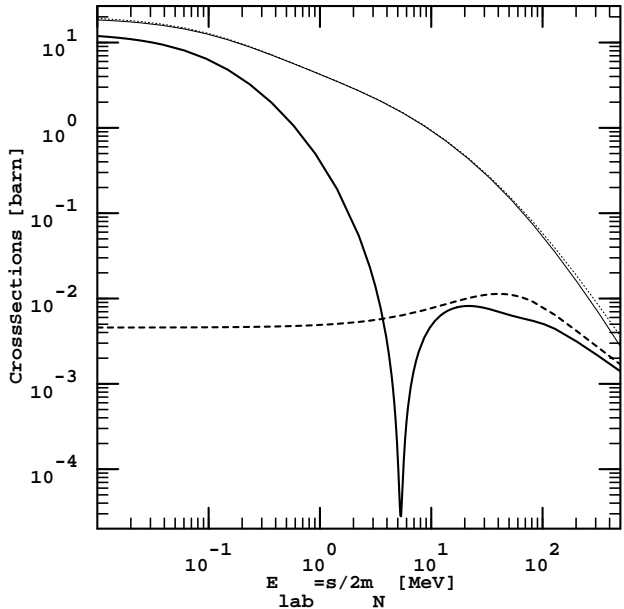}
\bigskip
\caption{The s-wave proton-neutron scattering cross section predicted
by the potential Eq.~(\ref{gauss}) (thin solid line), and measured (thin
dotted line) as a function of the laboratory kinetic energy
$E_{\rm lab}=s/2m_N$.
Also shown for this potential is the spin flip cross
section in Born approximation [Eq.~(\ref{sigmasfBorn}), thick dashed line]
and from phase shift analysis [Eq.~(\ref{sigmasfnum}), thick solid line].
The inverted resonance in the latter curve at $E_{\rm lab}/2\simeq2.2\,$MeV
stems from the deuteron bound state.}
\label{F1}
\end{figure}

For the proton-neutron interaction potential $V_S^{np}(r)$
for total spin $S$ we chose the following Gaussian potentials [such that
$U_\sigma^{np}=V_1^{np}-V_0^{np}$ and $U^{np}=(V_1^{np}+V_2^{np})/2$
in the notation of Eq.~(\ref{Vint})]:
\begin{eqnarray}
  V_0^{np}(r)&=&-33.6\,e^{-(r/1.77\,{\rm fm})^2}
               \,{\rm MeV}\nonumber\\
  V_1^{np}(r)&=&-84.7\,e^{-(r/1.36\,{\rm fm})^2}
               \,{\rm MeV}\,.\label{gauss}
\end{eqnarray}
Its strengths and ranges were fit to reproduce the experimental values
for the scattering lengths $a_S$ and effective ranges $r_{{\rm eff},S}$
which determine the low-energy expansion of the phase shifts
$\delta_{0,S}$~\cite{MS}:
\begin{equation}
  \cot\delta_{0,S}(s=4p^2)=-\frac{1}{pa_S}+\frac{p}{2}\,r_{{\rm eff},S}
  \,.\label{deltexp}
\end{equation}
As a result, the s-wave proton-neutron
scattering cross section predicted by Eq.~(\ref{gauss}) agrees with
the experimental one to within less than $5\%$ in the laboratory
energy range between 0 and $\simeq20\,$MeV (see Fig.~\ref{F1}).
In addition, the energy of the bound state resulting for $S=1$ coincides
with the deuteron binding energy within $5\%$. A central potential describes
the deuteron rather well since the contribution of the D state to the
bound state wave function is only about $6\%$.
Finally, we have compared the numbers for the weak neutral current
deuteron break up cross section resulting from Eq.~(\ref{sigmaD}) with
calculations in the literature~\cite{BKN}.
In the energy range between $\simeq5\,$MeV
and $40\,$MeV we found agreement to within about $10\%$.
This serves as a further check for the correct
normalization of our calculation.

\begin{figure}[ht]
\centering\leavevmode
\epsfxsize=3.2in
\epsfbox{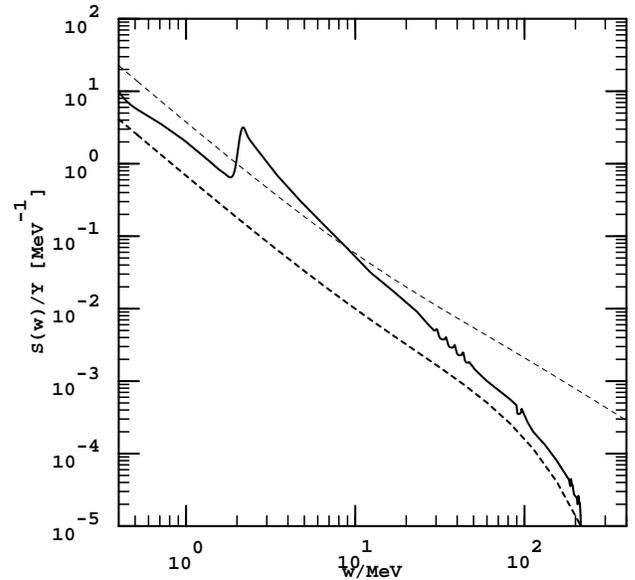}
\bigskip
\caption{The contribution to the dynamical nucleon spin-density
structure function $S_\sigma(\omega)/Y$ from proton-neutron scattering
as a function of $\omega$ in the long wavelength
limit $k\to0$. Shown is the Born approximation [Eq.~(\ref{SBorn}),
thick dashed line] and the result from phase shift analysis [Eq.~(\ref{S4}),
thick solid line] for the proton-neutron interaction potential
Eq.~(\ref{gauss}) and the estimate
Eq.~(\ref{quantum}) with Eqs.~(\ref{tildes}$-$\ref{GammaA}) for
an OPE potential in Born approximation (thin dashed line). The resonance
at $\omega\simeq2.2\,$MeV from the deuteron
binding energy is clearly visible in the thick solid line. The small wiggles
on this curve are caused by the finite numerical resolution of the
energy eigenvalues of the scattering states. For
$\omega<\Gamma_\sigma\simeq1.5\,$MeV, multiple scattering
effects start to become important and tend to suppress
$S_\sigma(\omega)/Y$ below the values shown here.}
\label{F2}
\end{figure}

Also shown in Fig.~\ref{F1} is the spin flip cross section as
calculated from the potential Eq.~({\ref{gauss}) both in Born
approximation [Eq.~(\ref{sigmasfBorn})]
and numerically from the phase shifts [Eq.~(\ref{sigmasfnum})]. It is
clearly seen that the Born approximation is far from being good.
The integrated spin flip rate Eq.~(\ref{Gammasf}),
$\Gamma_{sf}\simeq0.46Y_pY_n\,$MeV, differs by about a factor 2.5
from the Born approximation $\Gamma_{sf}^{\rm Born}\simeq0.21Y_pY_n\,$MeV.
This is also reflected by the nucleon spin-density structure function
calculated from Eqs.~(\ref{SBorn}) and (\ref{S4}) for $k\to0$, as shown in
Fig.~\ref{F2}. The phase shift analysis was performed by computing the
radial eigenfunctions up to some maximal orbital angular momentum
$l_{\rm max}=3$ above which they are close enough to the free
eigenfunctions to make a negligible contribution to Eq.~(\ref{S4}).
To achieve a sufficient resolution in the energy range of interest,
about 500 eigenfunctions had to be computed.
We verified that the resulting $S_\sigma(\omega)$
satisfies the f sum rule Eq.~(\ref{fsum2}) to within $10\%$. Note,
furthermore, that the Born approximation and the phase shift calculation
of the quantities shown in Figs.~\ref{F1}, and~\ref{F2} converge at
high energies where the second condition in Eq.~(\ref{Borncond}) is
asymptotically satisfied. We also verified that for a weak interaction
potential satisfying the first condition in Eq.~(\ref{Borncond}), the
Born approximation agrees well with the phase shift analysis over
the whole energy range as expected.

For comparison, Fig.~\ref{F2} also shows Eq.~(\ref{quantum}) with the
perturbative expressions Eqs.~(\ref{tildes}$-$\ref{GammaA}) for
the proton-neutron scattering contribution
to $S_\sigma$ based on the OPE potential (thin dashed curve).
After all, this curve reproduces the general normalization of the
spin-density structure function quite well, but it cannot reproduce
the quite prominent deuteron resonance. The non-vanishing pion mass
is taken into account in this curve and suppresses it by roughly
a factor 2 compared to calculations neglecting the pion mass.
Note the steepening at high $\omega$ of the curves for the potential
Eq.~(\ref{gauss}) in contrast to $S_{\sigma, {\rm OPE}}^{\rm Born}(\omega)$
which guarantees or violates f sum integrability, respectively.

The quantities of interest for neutrino diffusion as discussed in Sect.~II
are shown in Table~\ref{T1}. In calculating the cross section reduction
Eq.~(\ref{deltasigma}) we neglected the term $N_\sigma$ involving
spin-spin correlations. It is easy to see that the $S=1$ deuteron
bound state yields the contribution
\begin{equation}
  N_\sigma=\frac{C_{A,p}C_{A,n}}{C_A^2}\frac{Y_d}{3}\label{Nsigdeut}
\end{equation}
which implies a further reduction of $\left\langle\sigma_A\right\rangle$
because of the opposite sign of $C_{A,p}$ and $C_{A,n}$. This
corresponds to the fact that the cross section for elastic scattering
on deuterons is significantly smaller than that on free nucleons.
Also note that Eqs.~(\ref{Etrans}) and (\ref{recoil}) imply that
$\left\langle\Delta\varepsilon\right\rangle\propto T_\nu-T$ for
$|T_\nu-T|\ll T$. Finally, we observe that $\Gamma_\sigma/\Gamma_{sf}
\simeq\Gamma_\sigma^{\rm Born}/\Gamma_{sf}^{\rm Born}$, i.e. the
relation Eq.~(\ref{GammaBorn}) between the spin fluctuation rate
$\Gamma_\sigma$ appearing in $S_\sigma(\omega)$
and the spin flip rate $\Gamma_{sf}$ also holds beyond the Born
approximation.

\begin{table}[ht]
\begin{tabular}{lccc}
  & $\Gamma_\sigma/Y$ [MeV] & $\left\langle\Delta\varepsilon
    \right\rangle/(T_\nu-T)$
  & $\delta\left\langle\sigma_A\right\rangle/(\sigma_0 Y)$\\ \hline
  recoil & $-$ & 0.32 & $-$ \\
  OPE, Born & 3.6 & $0.18Y$ & $-0.14$ \\
  Eq.~(\ref{gauss}), Born & 0.66 & $0.029Y$ & $-0.018$ \\
  Eq.~(\ref{gauss}), exact & 1.5 & $0.15Y$ & $-0.076$ \\
 \end{tabular}
\bigskip
\caption{Comparison of the quantities shown in the top row for
the cases specified in the first column for proton-neutron interactions.
The first entry in the first
column is the proton-neutron potential used, and the second entry
indicates whether the Born approximation
[``Born'', according to Eq.~(\ref{SBorn})] or the exact two-nucleon
wave functions [``exact'', according to Eq.~(\ref{S4})] have been used.
For the OPE case, $C_p=1.09$ and $C_n=-0.91$ have been assumed.
Spin-spin correlations have been neglected in
$\delta\left\langle\sigma_A\right\rangle/\sigma_0$ (see text).}
\label{T1}
\end{table}

\begin{figure}[ht]
\centering\leavevmode
\epsfxsize=3.2in
\epsfbox{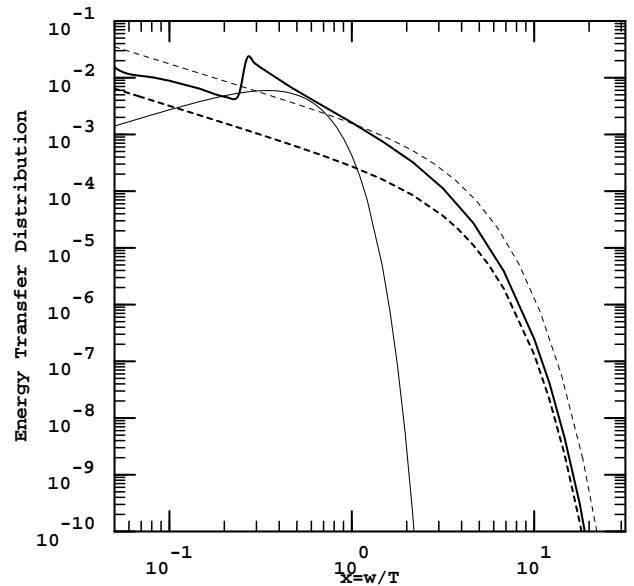}
\bigskip
\caption{The energy transfer distribution $S_\sigma(xT)
\left[1+\beta x/2+(\beta x)^2/12\right]$, resulting from proton-neutron
scattering, i.e. the integrand of
Eq.~(\ref{Etrans}) divided by $\omega$, as a function of $x=\omega/T$.
Here, the thick solid and dashed lines were obtained from the phase shift
analysis, Eq.~(\ref{S4}), and from the Born approximation,
Eq.~(\ref{SBorn}), respectively, for $Y=1$ and $k\to0$.
For comparison, the thin solid line
line was obtained from the recoil energy transfer, Eq.~(\ref{Srecoil}),
for a typical thermal momentum transfer $k=3T_\nu$, and the thin
dashed line is from Eq.~(\ref{quantum}) with
Eqs.~(\ref{tildes}$-$\ref{GammaA}) for an OPE potential
in Born approximation. For
$x<\Gamma_\sigma/T\simeq0.2$, multiple scattering
effects start to become important and tend to suppress the energy
transfer distribution below the values shown here.}
\label{F3}
\end{figure}

Clearly, the Born approximation for the potential Eq.~(\ref{gauss}) does
not give a good estimate to any of these quantities. However,
the numbers based on the OPE potential in Born
approximation give a reasonable estimate to most of the integrated
quantities in Table~\ref{T1}. In particular, these results confirm
that the average inelastic neutrino-nucleon energy transfer
$\langle\Delta\varepsilon\rangle$ is indeed comparable to the recoil energy 
$\langle\Delta\varepsilon\rangle_{\rm recoil}$, as suggested by
calculations employing the Born approximation for the OPE
potential~\cite{JKRS,RSS}. This energy transfer is, however,
differently distributed with a much longer tail to high energy transfers,
as can be seen in Fig.~\ref{F3}.

We have furthermore checked that the results for $S_\sigma$ calculated
from the phase shift analysis Eq.~(\ref{S4}) are insensitive to the
detailed shape of $U_\sigma^{np}(r)$ as long as it reproduces the
experimental phase shifts in the corresponding energy range. In particular,
properties of the potential at short distances $r$ influence
$S_\sigma(\omega)$ only for $\omega=p^2/m_N\gtrsim1/(m_N r^2)$.
For the conditions near the neutrinosphere it is therefore sufficient
that the potential reproduces nucleon-nucleon scattering up to
a few tens of MeV.

Towards the center of the hot neutron star, at densities around nuclear
density and $T\simeq30-50\,$MeV, predictions for the quantities shown
in Table~\ref{T1} by the OPE potential in Born approximation
are about 10 times higher than corresponding predictions based on
the potential Eq.~(\ref{gauss}) for which Born approximation and phase
shift analysis become rather similar. This shows that in this environment
weak interaction rates become quite sensitive to the short distance behavior
of the two-nucleon interaction potential which is different for these
two potentials. This can have important ramifications for neutrino
opacities and axion emissivities in the supernova core that are
usually based on these OPE calculations~\cite{FM,BT}. Whereas
calculations assuming an OPE potential should be a
reasonable approximation in the context of a ``cold'' neutron star this
is not necessarily the case for the much higher thermal energies
involved in a hot protoneutron star. Neutrino
opacities govern the cooling time
scale of the protoneutron star~\cite{KJR} while axion emissivities
determine axion mass bounds based on supernovae~\cite{KJSSTE}.
Apart from taking into account
many-body effects such as multiple scattering~\cite{RS2,JKRS,Sigl} a
more reliable calculation of these quantities thus requires to use
nuclear potentials that fit nucleon-nucleon scattering data also
at energies above a few tens of MeV to ensure the
correct small distance behavior. We leave that to a separate study.

\section{Summary and Conclusions}

We have discussed weak axial-vector current interactions involving
nucleons in hot non-degenerate nuclear matter at temperatures
around $10\,$MeV
and densities of a few percent of nuclear saturation density, i.e.
for conditions given in the vicinity of the neutrinosphere in a
type-II supernova. To describe such interactions in the limit of
non-relativistic nucleons we adopted the
structure function formalism for the nucleon spin-density.
Our special emphasize was on the energy transfer
between the weak ``probe'' and the nuclear medium which is
induced by the nucleon spin fluctuations caused by
the spin-dependent nucleon-nucleon interactions. To lowest order
in the nucleon-nucleon interactions, i.e. in Born approximation,
this is represented by nucleon bremsstrahlung. We have shown, however,
that the Born approximation is in general not a reliable estimate for these
effects. As an alternative, we have performed computations using
exact two-nucleon wave functions for a spherically symmetric two-nucleon
interaction potential that was fit to experimental data. In this
case, only proton-neutron interactions contribute to inelastic weak
neutral-current interactions with nucleons. We compared our calculations with
results for the corresponding contribution based on the usually
adopted OPE potential in Born approximation.
One gets rather good agreement for most integrated quantities such as the
reduction of the total neutrino scattering cross section and the
average energy transfer in a scattering event. In particular, we confirm
that the latter is comparable to the recoil energy, as suggested by
the results for the OPE potential in Born approximation. In contrast,
differential quantities such as the distribution of energy transfers
deviate significantly from predictions based on the Born approximation.
Our calculations specifically predict that the energy transfer peaks
at the deuteron binding energy. Resonances from higher nuclei which
have not been taken into account here such as helium are probably
less important for $T\lesssim10\,$MeV because they would appear at
higher energies and contribute less to the energy transfer
distribution (see Fig.~\ref{F3}).

The formalism presented here can be
extended to two-nucleon potentials that are not spherically
symmetric and to finite momentum transfer [see Eq.~(\ref{S3})].
Our results might have a significant
impact on the formation of neutrino spectra from type-II supernovae.
A quantitative understanding will, however, require detailed numerical
simulations. Finally, we demonstrated that weak interaction rates
in the hot supernova core are sensitive to the small distance behavior
of the nucleon-nucleon interaction potential which is not well
described by the usually adopted OPE potential.
This should be taken into account in future investigations.

\section*{Acknowledgments}
I am grateful to Georg Raffelt and Hans-Thomas Janka for many
discussions on this subject and for useful comments on the manuscript.
This work was supported, in part,
by the Deutsche Forschungs Gemeinschaft under grant SFB 375,
by the DoE, NSF, and NASA at the University of Chicago, and
by the DoE and by NASA through grant NAG 5-2788 at Fermilab.


\end{document}